\documentclass[a4paper,11pt]{article}
\pdfoutput=1 

\usepackage{jcappub} 

\usepackage[T1]{fontenc} 

\title{\boldmath Powerful flare phenomena in water vapor maser lines in the emerging protostellar system with protoplanetary disks IRAS 16293-2422}

\author[a,1]{A.E. Volvach,\note{Corresponding author.}}
\author[a]{L.N. Volvach,}
\author[b]{M.G. Larionov}

\affiliation[a]{Radio Astronomy and Geodynamics Department of Crimean Astrophysical Observatory RAS, Katsively, RT-22 Crimea}
\affiliation[b]{Astro Space Center, Lebedev Physical Institute, Russian Academy of Sciences, Profsoyuznaya ul. 84/32, Moscow, Russia}

\emailAdd{volvach@meta.ua}

\abstract{Based on the long-term monitoring data of the water maser since 2019.0 to 2021.0 allowed us detecting in IRAS 16293-2422 two powerful phenomena lasted about year at radial velocities near 6 and 8 km~$s^{-1}$.  
In both cases, powerful short flares were located on the tops of less powerful, but more prolonged ones (2.5 and 0.5 kJy), radiation of which initiated the emission of more powerful flares. 
For the first time, configurations of several emitting maser spots located at the line of sight to the observer were discovered experimentally.  
This made it possible to confirm the hypothesis of activation of the water maser, based on an increase in the amplification length of the maser due to several maser condensations located at the line of sight to the observer. 
The unsaturated state of the most powerful and shortest maser flares, as well as the saturated state of the weak, has been established. 
New important parameters of the water maser and the assumed location of the maser spots have been obtained.
}

\begin{document}
\maketitle
\flushbottom

\section{Introduction}
\label{sec:intro}

It has long been known that stars of both early and late spectral types in 75\% of cases form binary and multiple systems. 
Stars often form together with companions \cite{1,2}. 
They form from collapsing cores that are the densest parts of a large molecular cloud, in which revolution, turbulence and magnetic fields cause elliptical symmetry in a system \cite{3,4,5}. 
The IRAS 16293-2422 (IRAS 16293 hereafter) is protostellar objects of Class 0, as seen in surveys of young stellar objects \cite{6,7}. 
This young system was discovered more than a third of a century ago and has been intensively researched in various wavelength ranges since then \cite{8,9}.
IRAS 16293 is located in the $\rho$ Ophiuchus cloud complex at a distance of $\sim$140 pc \cite{10} and has a complex structure on a scale of about 3000 AU. 
This system has a gas and dust disks of 12 and another less than 3.6 AU, located around low-mass protostars of solar masses and moving, possibly in Kepler's orbits \cite{11}. 
IRAS 16293A is highly active with two different directions bipolar outflows connected with protostellar system \cite{12,13,14,15}.

Submillimeter interferometers made it possible to investigate the internal structure of protostars, including IRAS 16293 \cite{16,17,18,19}.
Near IRAS 16293, two sources of the continuum A and B were discovered, separated by of $\sim$700 au \cite{20}. 
VLA observations at $\lambda$2 cm were able to resolve in A two compact sources A1 and A2 separated by a distance of $\sim$50 AU \cite{20}. 
A third companion (B) was also installed, located about 800 AU from A1 and A2.
Thus, IRAS 16293 is a triple protostellar system, gravitationally bound, parts of which revolve around a common center of gravity. 
The period of mutual circulation of A1 and A2 is about 320 years, while A and B are several thousand years \cite{11}.

Interferometric observations have also established that there are groups of maser condensations (spots) associated both with the outflow and, possibly, located in gas and dust disks around the protostars \cite{20,21,22}. 
To determine the location of water masers, simultaneous monitoring observations in the single radio telescopes and interferometric modes are required. 
It is rather difficult to make such observations. 
Now, we can only assume that IRAS 16293 is double protostellar nature with a mass of 2-3 $M_{\odot}$ \cite{20,23,24,25}.
ALMA observations with a resolution of 6.5 AU, revealed two compact dust thermal emission sources, coincident with the location of the sources A1 and A2, which confirms them as a binary protostars system \cite{20}. 
It may be one of the disks where our water vapor maser are also located.

The paper presents new data on powerful long-term water maser phenomena in features with a radial velocities near 6 and 8 km~$s^{-1}$, occurred in the young binary system IRAS 16293 and an interprenation of data obtained. 

\section{Observation and data processing}
\label{sec:obser}

The observations at a frequency 22.235 GHz of the $6_{16} - 5_{23}$ water-vapor maser transition  have been made from 2019 January to 2020 February. 
The 22-m Simeiz telescope (RT-22) and a spectral-polarimetry radiometer with a parallel Fourier spectrum analyzer were used. 
Resolutions of the spectrum analyzer were 8 and 2 kHz (0.105 and 0.03 km~$s^{-1}$), respectively \cite{26}. 
The half-width of the radiation pattern of the radio telescope (FWHM) was 2.5-arc min, sensitivity - 13 Jy\,K$^{-1}$. 
Depending on the weather condition, the system temperature varied in the range 120 to 150 K. 
The collected spectral data were corrected for atmospheric absorption and changes in the effective areas of the radio telescope at different elevation angles.

\section{Results}
\label{sec:res}

Observations of the water vapor maser near 6 and 8 km~$s^{-1}$ features in IRAS 16293 are shown in Fig. 1. 
Intervals between data collection were 1-2 days. 
Flares at both 6 and 8 km~$s^{-1}$ occurred overlapping each other in time. 
All flares for the first event at 6 km~$s^{-1}$ lasted about six months, for the second at 8 km~$s^{-1}$ - about two months. 
The data obtained indicate a complex distribution of emitting spots in the maser cluster. 
This is especially observed for a feature near 6 km~$s^{-1}$. 
Each of these complex flare phenomena consist of several shorter flares. 

To understand the physical nature of these flare conglomerates, we use a time-spectral research method. 
It is based on the analysis of flux density monitoring data and spectral data simultaneously. 
In Fig. 1, a flare phenomenon of about 6 km~$s^{-1}$ is shown in blue and about 8 km~$s^{-1}$  - in red. 
In Fig.1 flares of the first and second phenomena are numbered.
The figure also shows less powerful flares within 1 kJy, which are present for both at feature 6 and 8 km~$s^{-1}$. 
In this work, we will not consider them, since this is not included in the topic of our consideration.
We have conventionally divided the first flare event into six shorter flares, and the second into three. 
For now, let's start our consideration with the ones indicated in Fig. 1 flares.

\begin{figure}[tbp]
\centering 
\includegraphics[width=3.5in,height=2.5in]{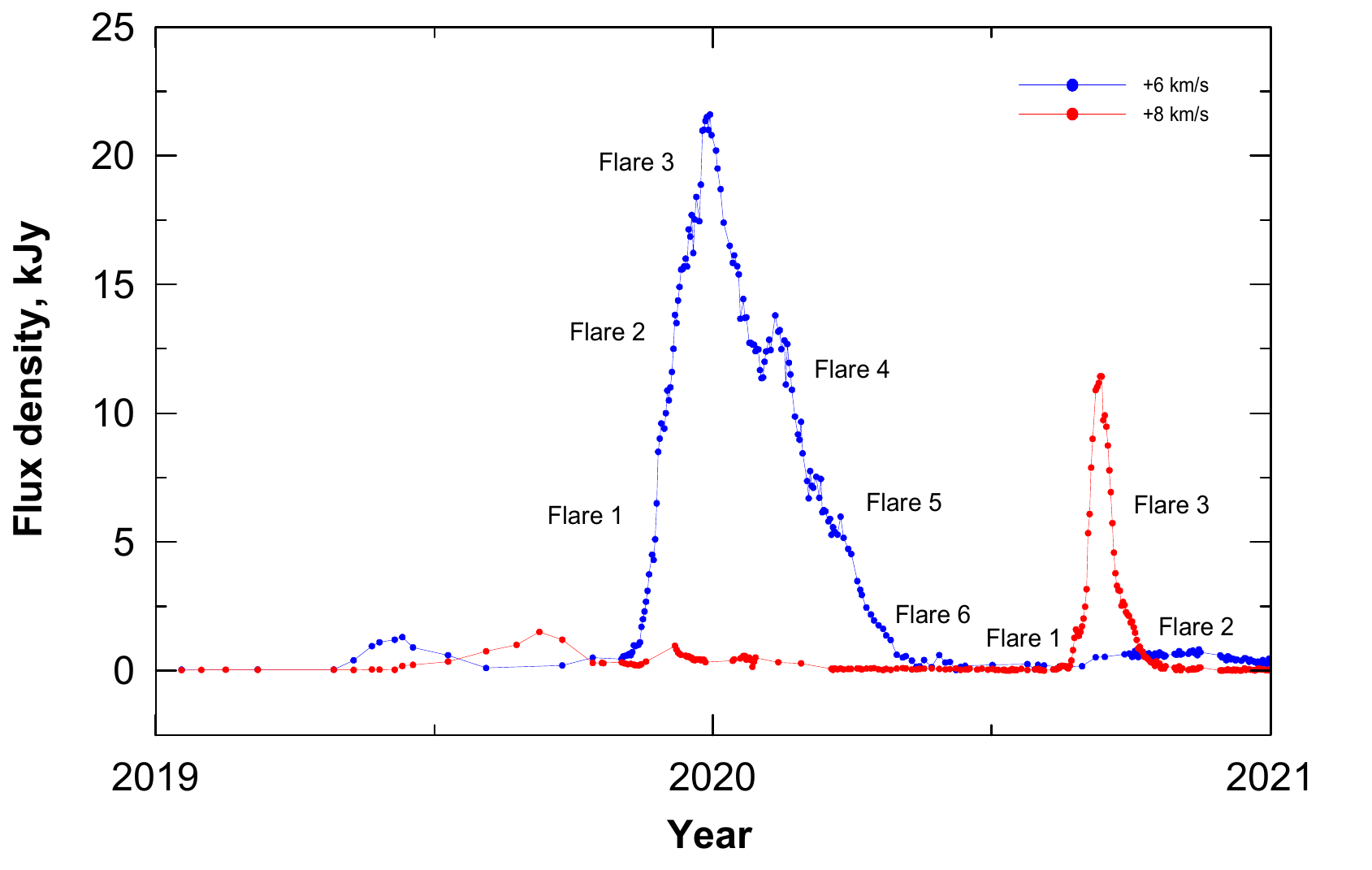}
\caption{\label{fig:i} 
The powerful flare phenomena that occurred in IRAS 16293 at near 6 km~$s^{-1}$ are shown in blue and at near 8 km~$s^{-1}$ in red.}
\end{figure}

The powerful flare phenomenon at feature 6 km~$s^{-1}$ began with a sharp increase in the radiation flux density. 
The duration of the rise in the flux density of flare 1 (so we call it) is about 28 days (14 days at the level of half of the maximum flux density, Fig. 1). 
It is known from interferometric data that some maser sources  have sizes in the range of 0.5-2 AU (in Orion KL, for example). 
In fact, the size of the maser spots may be smaller, since there is a scattering effect in the interstellar medium and in the medium in which the source itself is located \cite{27}. 
If the flare flux density increases exponentially, then the dependence of the line-width on the flux density will be the same as shown in Fig. 2 for flares at features near 6 and 8 km~$s^{-1}$ (first and second phenomena). 
Exponential grow and fall of flux density during a flare together with a decrease in line width with grow flux density are predicted for a maser in an unsaturated state.
We got the best-fit line (Fig. 2) and the relation describes the experimental dependence \cite{28}:

\begin{equation}
\label{eq:x}
\begin{split}
\frac{1}{(\Delta v)^{2}} = a + b \cdot \ln S.
\end{split}
\end{equation}

where  $\Delta v$ - a line half-width (a full width at half maximum), $S$ - a flux density. 
The calculated error $\Delta v$ does not exceed the velocity resolution that we obtained in each observation (from 0.03 to 0.1 km~$s^{-1}$).
In that equation $a$, $b$ are coefficients. 

\begin{figure}[tbp]
\centering 
\includegraphics[width=.49\textwidth,origin=c,angle=0]{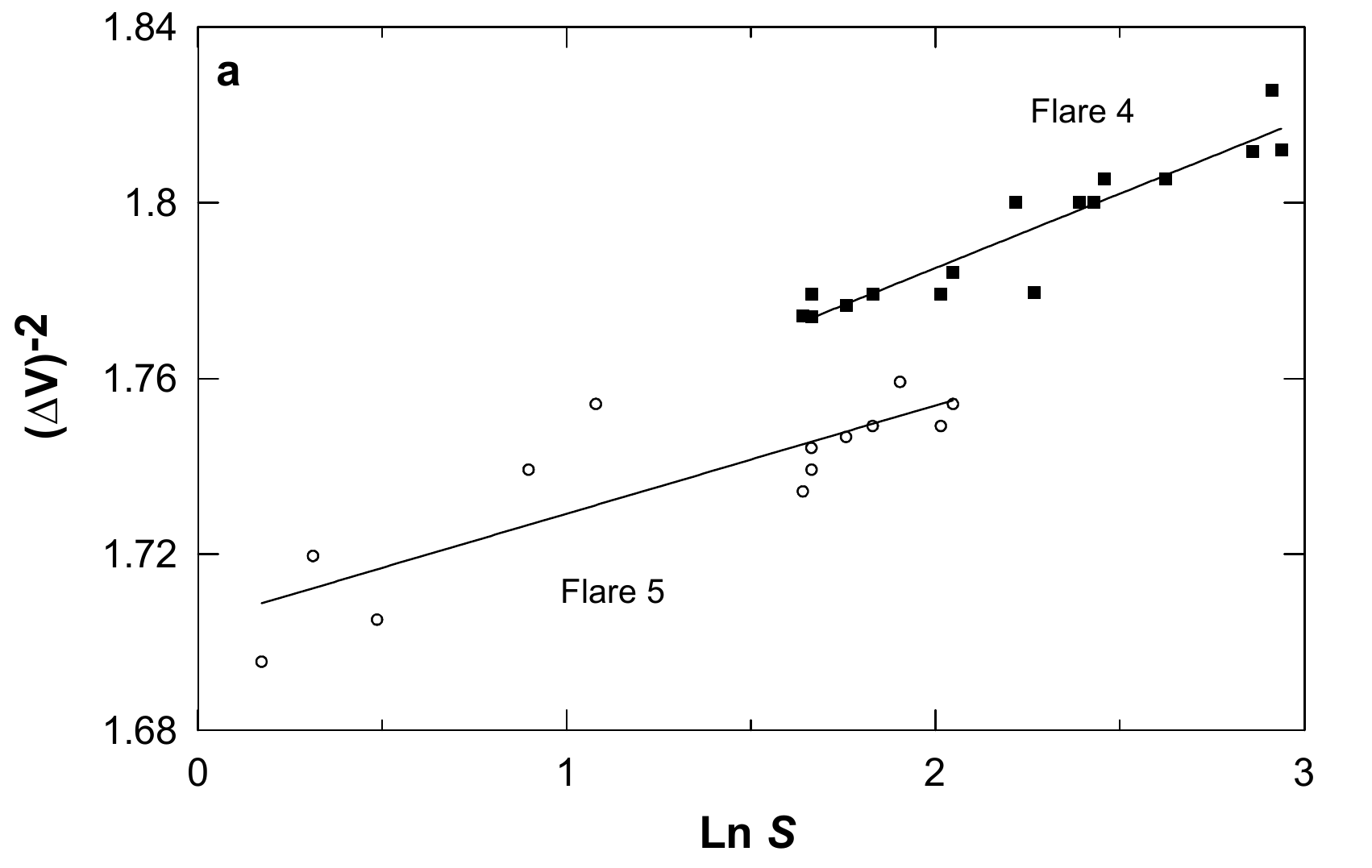}
\hfill
\includegraphics[width=.49\textwidth,origin=c,angle=0]{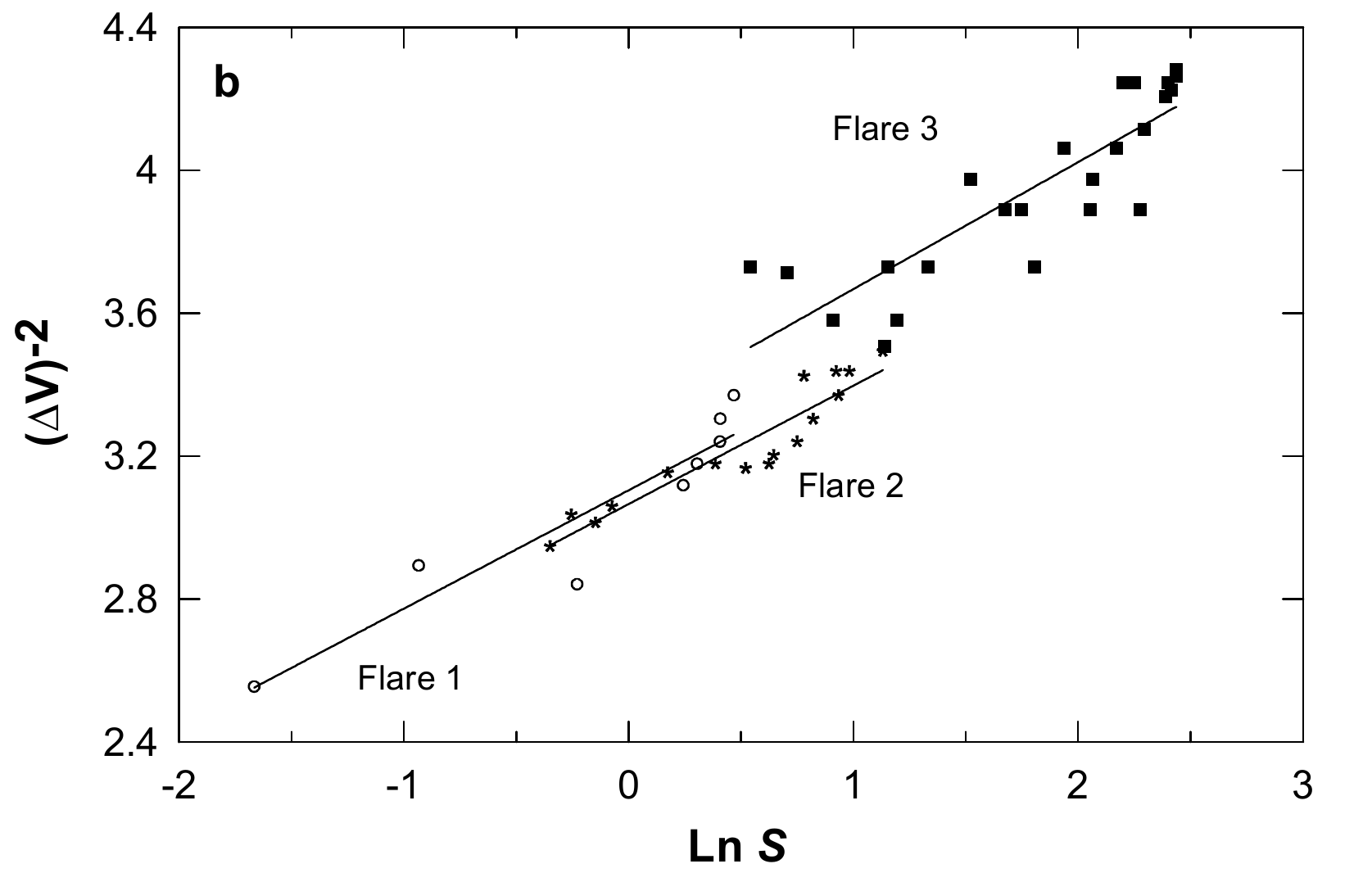}
\caption{\label{fig:i} 
The dependence between the spectral linewidth in the power of minus two ($\Delta v)^{-2}$ km~$s^{-1}$ and the natural logarithm of the flux density ln(S) kJy in IRAS 16293: (a) for Flare 4, 5 at the feature near 6 km~$s^{-1}$ and (b) for Flare 1, 2, 3 at the feature about 8 km~$s^{-1}$.
}
\end{figure}

Table 1, 2 lists the parameters of the flares for both flare phenomena.

\begin{table}[tbp]
\begin{center}
\caption{\label{tab:i} Characteristics of flares at the feature near 6 km s$^{-1}$ in IRAS 16293.}             
\begin{tabular}{|c|c|c|c|c|c|}
  \hline
Flare, features	        & $S_{max}$, & $\Delta v_{measure}$  & $T_{k}$, K	& $\Delta v_{model}$,	& $T_{k}$, K \\
$\sim$ 6 km s$^{-1}$	&   kJy	    & at S$_{max}$           & measure       &  at S$_{max}$          & model \\
  \hline
1	& 9.5  & 0.63 & 150 & 0.75  & 245 \\
  \hline
2	& 16.0 & 0.62 & 146 & 0.75  & 242 \\
  \hline
3	& 22.0 & 0.63 & 150 & 0.64  & 178 \\
  \hline
4	& 13.5 & 0.63 & 150 & 0.74  & 238 \\
  \hline
5	& 7.5 & 0.62 & 146  & 0.75  & 245 \\
  \hline
6	& 5.0 & 0.64 & 155  & 0.75  & 245 \\                
  \hline
7	& 2.5 & 2.8   & 2980 &        & \\                
  \hline
\end{tabular}
\end{center}
\end{table}
\begin{table}[tbp]
\begin{center}
\caption{\label{tab:i} Characteristics of flares at the feature near 8 km s$^{-1}$ in IRAS 16293.}             
\begin{tabular}{|c|c|c|c|c|c|c|}
  \hline
Flare, features	        & $S_{max}$, & $\Delta v_{measure}$  & $T_{k}$, K	  & $\Delta v_{model}$,   & $T_{k}$, K & $V_{max}$, \\
$\sim$ 8 km s$^{-1}$	&   kJy	    & at S$_{max}$            & measure       &  at S$_{max}$          & model        & km s$^{-1}$ \\
  \hline
1	& 1.8   & 0.59 & 132  & 0.55  & 115   & 7.78 \\
  \hline
2	& 2.5   & 0.61 & 141  & 0.56  & 119   & 7.77 \\
\hline
3       & 11.5 & 0.60 & 139  & 0.50  & 95   & 7.76 \\
  \hline
4       & 0.5   & 2.8  & 2980 &        &        & 7.61 \\
  \hline
\end{tabular} \\
\end{center}
\small {Notes to the Table 1, 2: 
$\Delta v_{measure}$ (km~$s^{-1}$) - the line half-width obtained from observations, $\Delta v_{model}$ the line half-width obtained from model dependence, $S_{max}$ (kJy) is the maximum value of the flare flux density, $T_{k~(measure)}$ - kinetic temperature H$_2$O, obtained from the observational data ($\Delta v_{measure}$) and $T_{k~(model)}$ - kinetic temperature H$_2$O, obtained by the used model ($\Delta v_{model}$), were calculated according to \cite{29}, $V_{max}$ - the frequency (the velocity) of the flare at the maximum flux density, $k$ is Boltzmann's constant,
$\frac{\sum}{n}$ and $\sigma$ - measure and root mean square error (about 10\% of the value). 
The calculated error $\Delta v$ does not exceed the velocity resolution (from 0.03 to 0.1 km s$^{-1})$
}
\end{table}   

From the analysis of the parameters (Table 1, 2), new important conclusion can be drawn. 
Flares in both flare events (near 6 and 8 km~$s^{-1}$) are well approximated by the linear dependences presented for unsaturated masers (Fig. 2). 
The line widths measured directly from the observational spectra and determined from the model dependences coincide within the observational errors. 
Less powerful flares in both cases (Flares 7 for the phenomenon near 6 km~$s^{-1}$ and flare 4 for 8 km~$s^{-1}$) have uncondensed spectral lines, which means they most likely belong to saturated masers. 
In some cases, the degree of line narrowing in unsaturated masers is 4.75 times. 
This is also an important result, as it shows that the detected narrowing is close to the limit (5-6 times) \cite{30}. 
The kinetic temperatures in both cases are close to 3000 K, which provides a high degree of ionization and the necessary population of signal levels in the rotational transitions of the water molecule. 
This can be considered one of the most important results obtained. 

An ultra-short, sufficiently powerful flare 1 (about 2 kJy) with a duration of about 4-5 days was detected near 8 km s$^{-1}$ (Fig.1). 
This is about ten times less than the duration of Flare 2 of the same flare phenomenon. 
The maser spots responsible for these flares are most likely located in the same maser cluster, which means that their sizes can be compared. 
They also differ tenfold. 
This is the first reliable case when it was possible to register identical emissions from maser formations, that differ in size by a factor of 10 (10$^3$ in volume). 
The conclusion is not trivial. 
How can this be? 
After all, the agent initiating the maser emission at a given time interval is the same. 

A powerful phenomena are discovered represents superposition of individual flares of maser radiation. 
In the case of the first phenomenon near 6 km s$^{-1}$, its onset is accompanied by a sharp increase in the spectral radiation flux density. 
The flux density of three first flares increases exponentially. 
The flares are partially overlapping each other in time. 
The drop in the flux densities of flares 4, 5, 6 also occurs exponentially. 
The spectra of the feature near 6 km s$^{-1}$ for the flare 1 and 4, for example, are shown in Fig. 3.
 
It can be noted, that in addition to the six flares, there is one more (Flare 7), less powerful, but long lasting, the line half width of which is several times larger than for all six flares. Its amplitude is in the range of 2.0-3.0 kJy in time of action each of six flares. 
Flare seven more visible at the lower values of the flux densities. 
At the top of this flare, all the other six flares appear. 
Flare 7 is the base for all the others, creating about 2.5 kJy input flux for them. The water masers of six short, powerful flares appear to be in an unsaturated state, while the long flare maser 7 is in a saturated state. 
The state of the flare 7 maser is confirmed by the presence of a broad line that does not narrow as in the case of an unsaturated maser. 
The situation is being implemented, when powerful flares of the water maser occur due to an increase in the pumping length of the maser during cascade amplification.

\begin{figure}[tbp]
\centering 
\includegraphics[width=.49\textwidth,origin=c,angle=0]{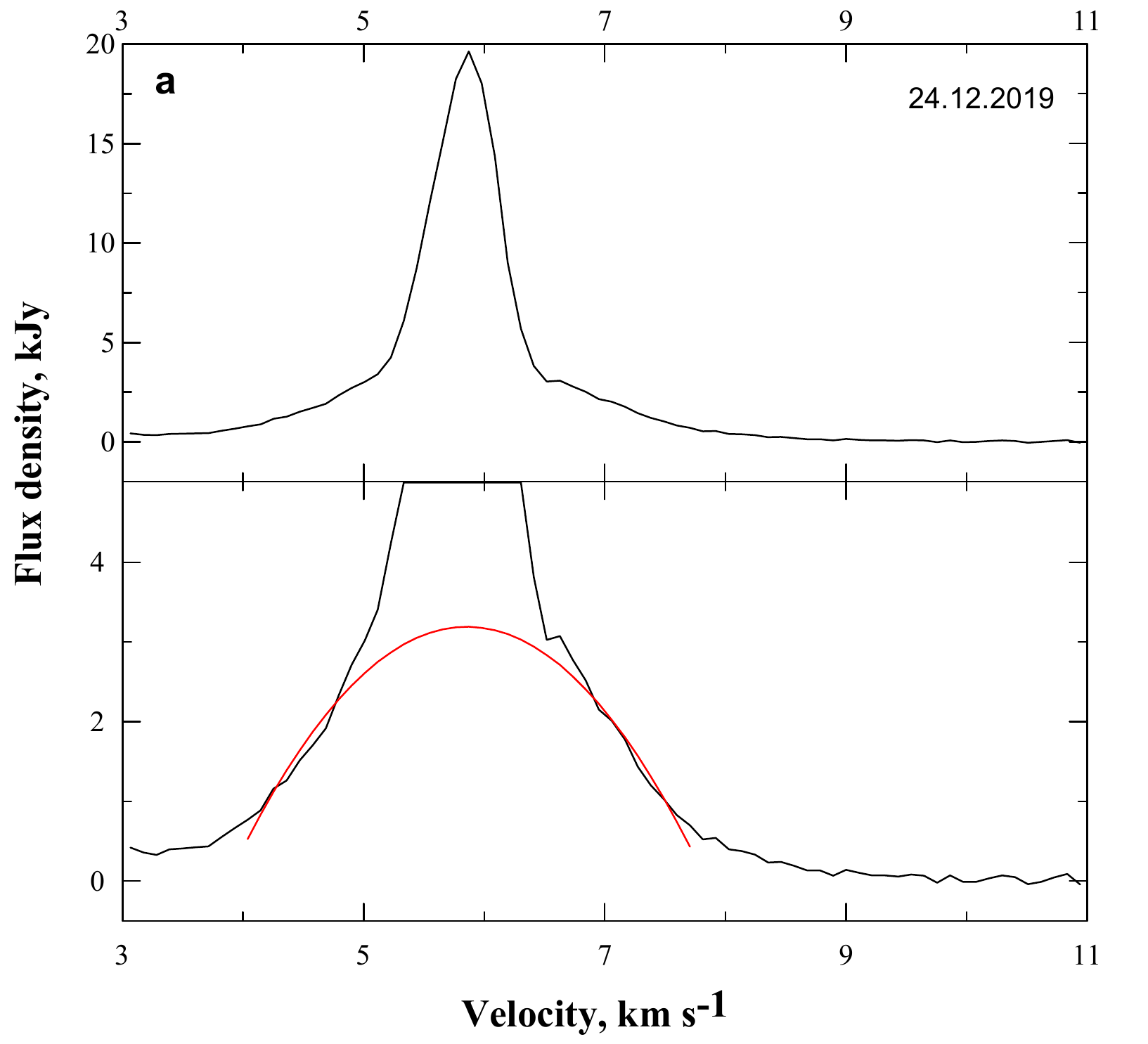}
\hfill
\includegraphics[width=.49\textwidth,origin=c,angle=0]{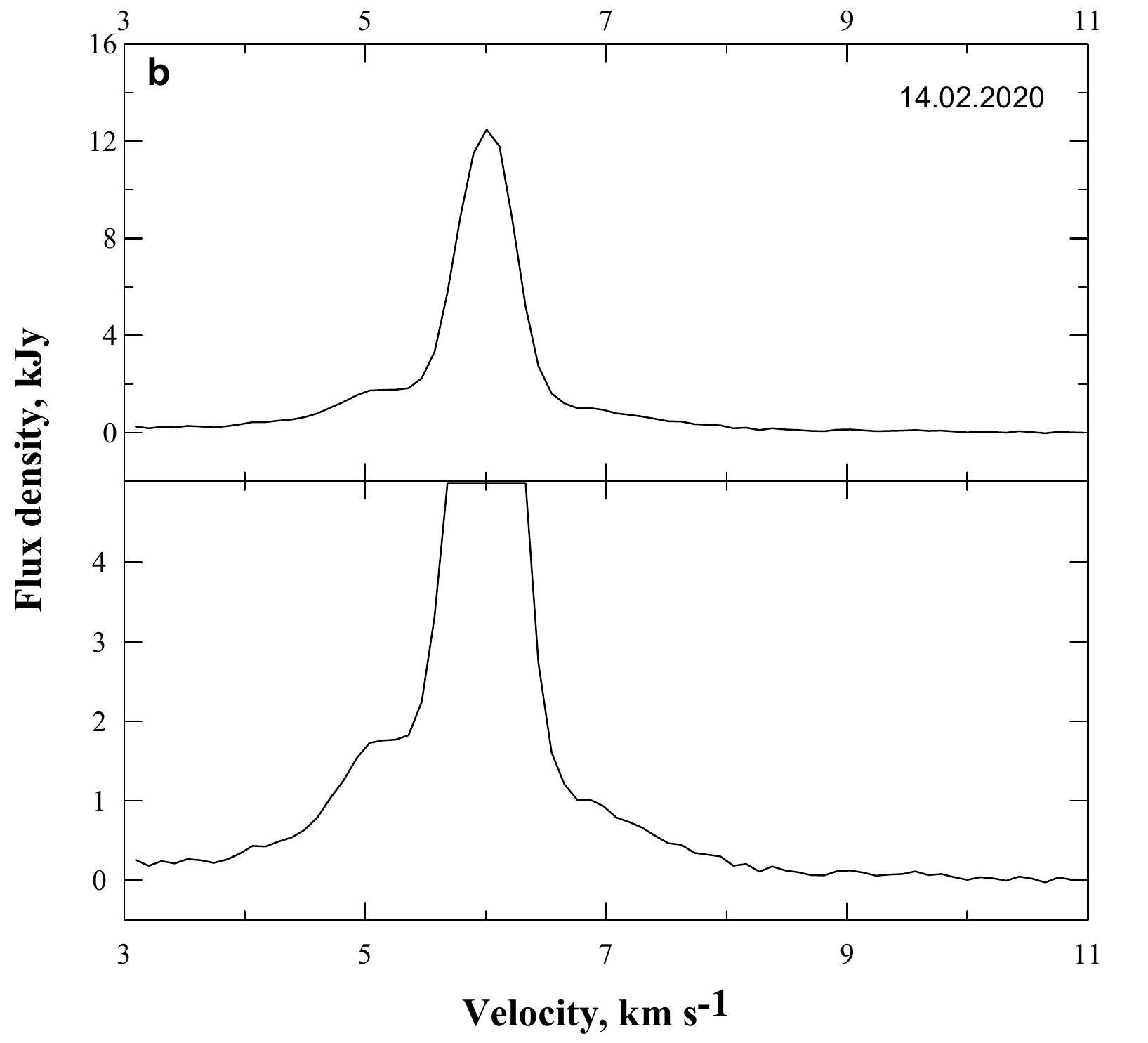}
\caption{\label{fig:i} 
IRAS 16293. The spectrum of the feature near 6 km~$s^{-1}$: 
a) for the Flare 3 (at the maximum flux density Flare 7, $S_{max}$ $\approx$ 3 kJy), 
b) for the Flare 4 (at close to the minimum flux density Flare 7, $S_{min}$ $\approx$ 2 kJy).
}
\end{figure} 

In the second phenomenon near 8 km~$s^{-1}$, super-short Flare 1 started at first with a duration of about four days at the level of half the maximum flux density 1.8 kJy (Fig.1). 
The Flare 2, with a duration of 40 days and the amplitude of about 2.5 kJy is accompanied by a more powerful Flare 3 with a maximum flux density $S_{max}$ of 11.5 kJy.
The duration this most powerful flare, is also very insignificant for about 15 days at the level of half the maximum flux density $S_{max}$.

The spectrum of the features near 8 km~$s^{-1}$ for the flare 1, 2 are shown Fig. 4.

\begin{figure}[tbp]
\centering 
\includegraphics[width=.49\textwidth,origin=c,angle=0]{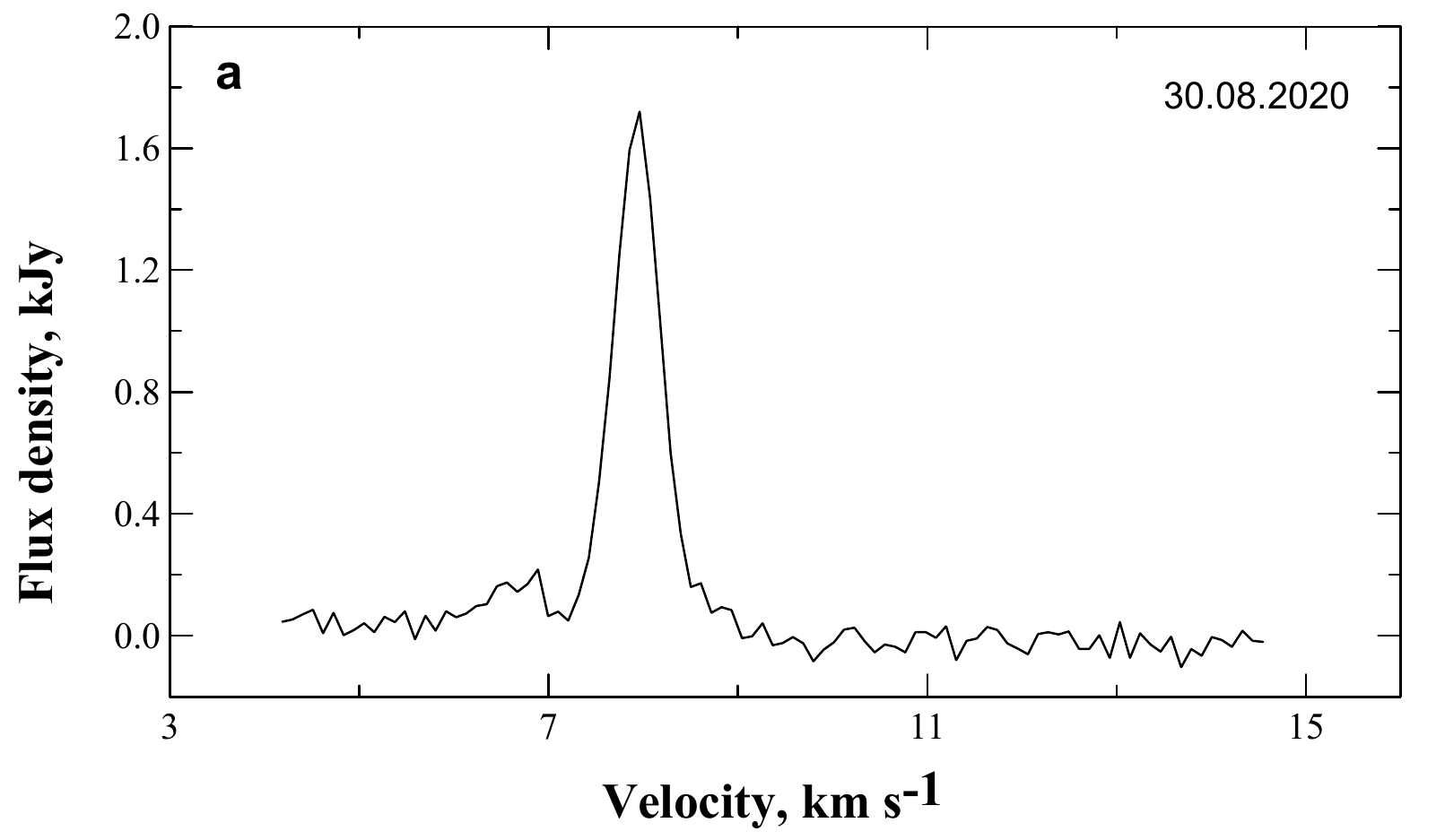}
\hfill
\includegraphics[width=.49\textwidth,origin=c,angle=0]{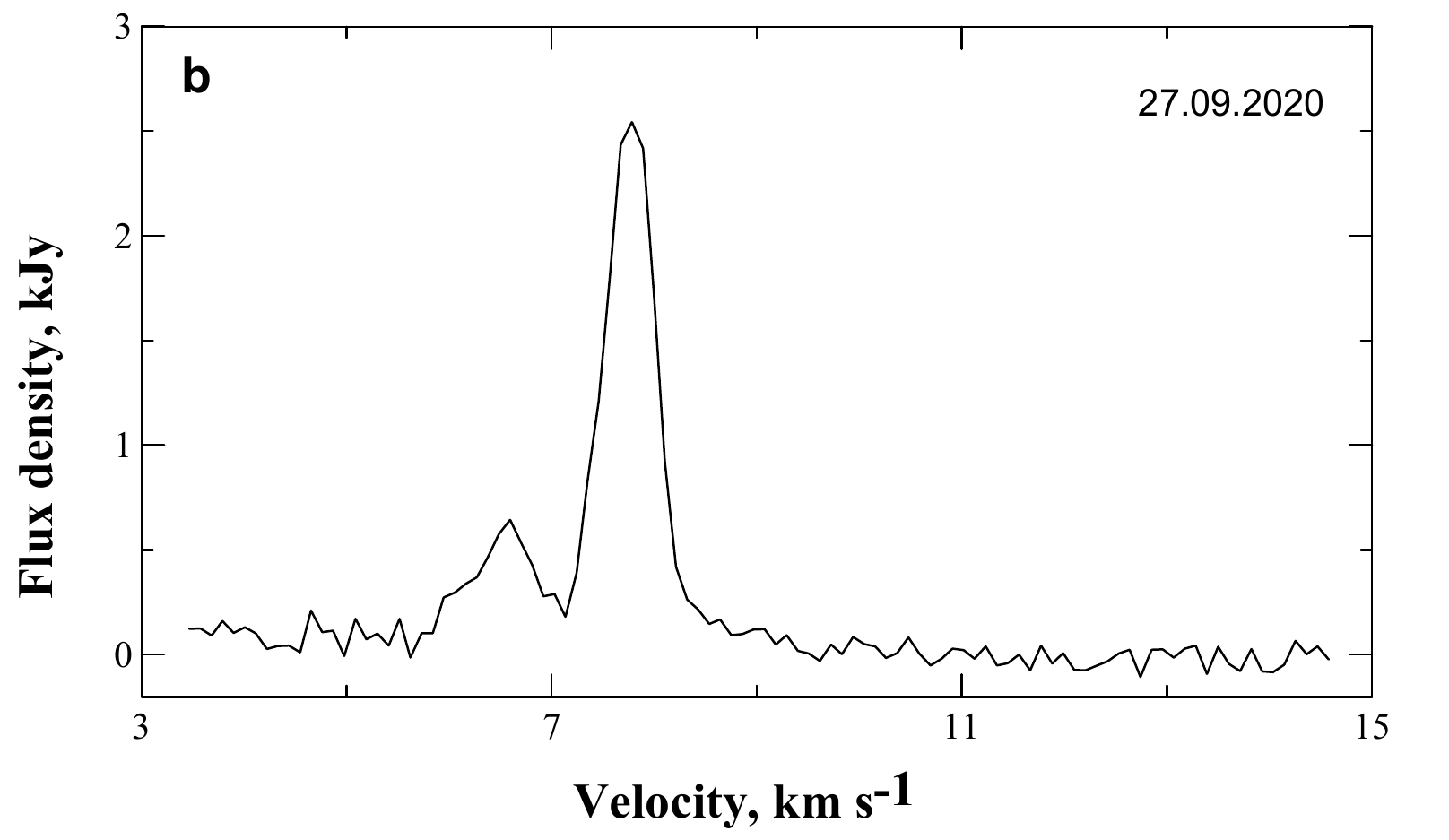}
\caption{\label{fig:i} 
IRAS 16293. The spectrum of the feature near 8 km~$s^{-1}$: a) for the Flare 1, b) for the Flare 2.
}
\end{figure} 

As in the case of the first flare phenomenon near 6 km s, the second flare phenomenon have of a less powerful flare 4 with an amplitude of about 0.5 kJy (Fig. 5).

\begin{figure}[tbp]
\centering 
\includegraphics[width=.49\textwidth,origin=c,angle=0]{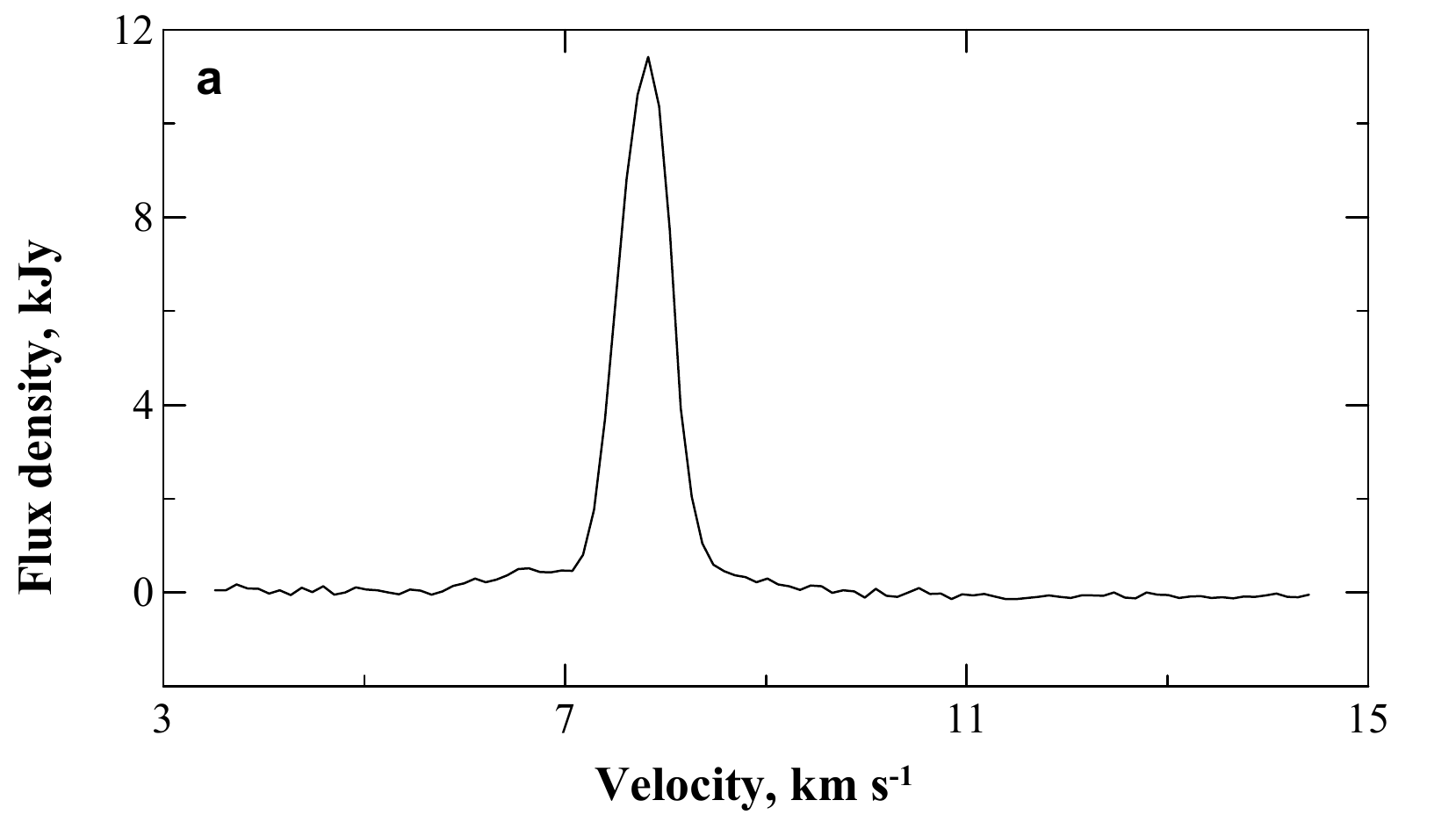}
\hfill
\includegraphics[width=.49\textwidth,origin=c,angle=0]{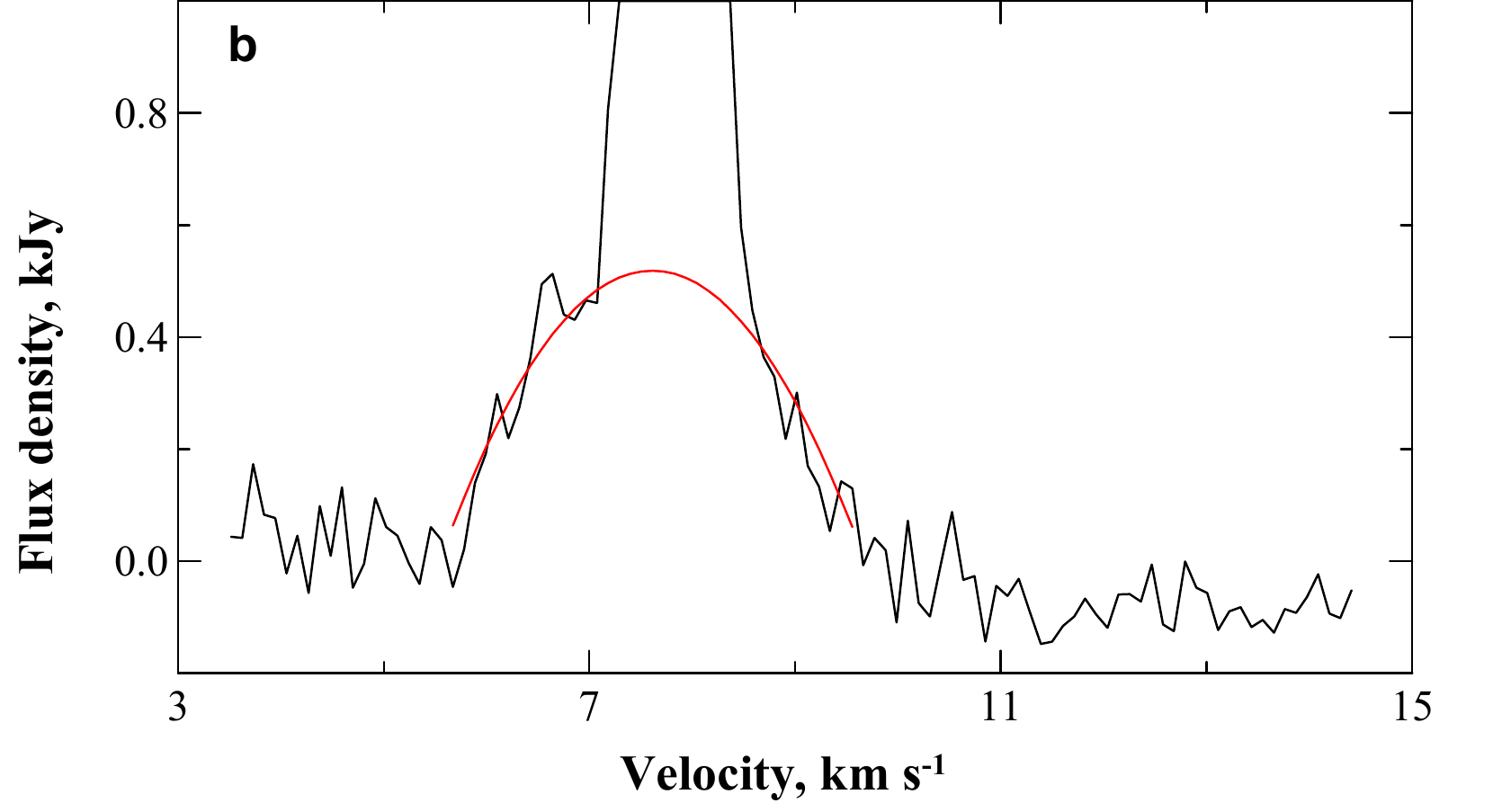}
\caption{\label{fig:i} 
The spectrum of the water maser feature in IRAS 16293 for the Flare 3 and 4 (12.09.2020). 
The maximum amplitudes 0.5 kJy and localization 7.61 km~$s^{-1}$ for Flare 4 is obtained by fitting a Gaussian curve (red line) into the spectra of the line and finding the minimum of the sum of squares of the deviations between them. 
The localization 7.76 km~$s^{-1}$ for the Flare 3 obtained also by fitting a Gaussian curve (red line) into the spectra of the line.
}
\end{figure} 

The Flare 4 is fundamental for the other three flares. 
Flare 4 becomes noticeable at low values of flux densities (the scale of the ordinate axis in Fig. 5 was increased by an about order of magnitude). 
The center frequencies of all three lines (Fflares 1, 2, 3) are within 0.1 km~$s^{-1}$ (Table 2).

\section{Discussion}
\label{sec:dis}

Two powerful flare phenomena that we detected in IRAS16293 at features near 6 and 8 km~$s^{-1}$ are formed by a set of maser spots to the observer's line of sight. 
In each case we see, that there are two types of flares with different states of the water maser - saturated and unsaturated 
ones. 
Saturated masers always play the impottant role in the formation of unsaturated masers and, by their radiation, initiate the emission of unsaturated masers. 
The natural assumption arose that all powerful short water maser flares were "constructed" in this way. 
But the situation can be even more complicated. In the second flare phenomenon near 8 km s, a powerful short Flare 3 "starts" from the top of 
an ultra-short Flare 1, the maser of which is in an unsaturated state (Fig. 1). 
At the same time, Flare 4 with a saturated maser also "feeds" powerful Flare 3 and Flares 1, 2 by its radiation (Fig. 5). 
All three flares of unsaturated masers with close frequencies can occur in maser spots of the same cluster.
This explains why Flare 3 is so powerful. 
Due to the rarity of such phenomena, which were registered in detail, we can't yet unambiguously confirm these assumptions. 
Therefore, any new flare phenomenon, recorded and analyzed in detail, is of great interest for solving the problem of water maser radiation. 

A deeper analysis of the data for features near 6 km~$s^{-1}$ shows that there are two groups of maser spots with a detectable difference in radial velocities (Table 3). 
The first group of Flares (1, 2, 3) has similar radial velocities with an average value of 5.88 km~$s^{-1}$. 
The second group (4, 5, 6) also has close radial velocities with an average value of 6.05 km~$s^{-1}$. 
The difference in radial velocities between these two groups is about 8 rms, which is certainly a very significant result that needs to be understood and explained. 
One possible explanation is that the maser spots are grouped into two clusters, which are involved in motions along Keplerian orbits. 
In this case, maser spots (Flares 1-6) can be located in front of the maser spot responsible for flare 7, which initiates the appearance of six powerful flares. 
The change in the radial velocity of Flare 7 (0.06 km~$s^{-1}$) also occurs, but to a much lesser extent, which may be natural if this maser 
condensation is in a different orbit with an another radial velocity (Table 3). 
Thus, the results obtained can indirectly indicate the orbital motions of maser spots around a certain center of gravity.

\begin{table}[tbp]
\begin{center}
\caption{\label{tab:i} Values of the central velocities of lines of Flares 1-6 near 6 km~$s^{-1}$ and velocities of line 7 during the passage of each of Flares 1-6.}   
\begin{tabular}{|c|c|c||c|c|c|c|}
  \hline
Flare  & $v_{(central)}$, km s$^{-1}$  & $v_{Flare7}$, km s$^{-1}$ &  Flare  & $v_{(central)}$, km s$^{-1}$  & $v_{Flare7}$, km s$^{-1}$ \\
  \hline
1	& 5.90  &  5.95 & 4	& 6.05  &  5.87   \\
\hline
2	& 5.86  &  5.96  & 5	& 6.07  &  5.87  \\
  \hline
3	& 5.87  &  5.92  & 6	& 6.03  &  5.90  \\
  \hline
$\frac{\sum_{}^{}}{n}, \sigma$ & \bf{5.88 $\pm$ 0.02} & \bf{5.94 $\pm$ 0.02} & $\frac{\sum_{}^{}}{n}, \sigma$ & \bf{6.05 $\pm$ 0.02} & \bf{5.88 $\pm$ 0.02} \\
  \hline
\end{tabular} \\
\end{center}
\small {Notes: $v_{central}$ is the flare frequency in units of km s$^{-1}$, $v_{Flare 7}$ - the flare frequency 7 at the maximum level of each of the flares in units of km s$^{-1}$,  $\frac{\sum_{}^{}}{n}, \sigma$ - the mean values and root mean square errors. 
The errors in the radial line velocities were determined by the value of the spectral resolution equal to 0.03 km s$^{-1}$.}
\end{table}   

Powerful flares of the water maser is of great importance in understanding the physical processes of maser generation and would provide a strong constraint on the possible model of radiation generation. 
Maser pumping depends on the primary scheme pumping, the type of energy sink of the maser heat engine (both such as collisional and radiative), and the type of transition that makes up the pump cycle (rotational or vibrational). 
The powerful flares of  water maser are explained in this framework by collision--collision pumping mechanism in rotational levels ($CCr$)   because of certain difficulties are encountered with other pumping mechanisms, in particular, the radiative ones \cite{31,32}. 
The mechanism of CCr maser pumping behind the shock front was proposed \cite{30,33}; where there are supersonic gas outflows from young star and their interactions with surrounding gas clouds. 
It produces supersonic shock waves required for maser pumping. 
%

In the case of class 0 protostars, the question of the sources of primary energy release that ensures the pumping of powerful water masers remains open.
This situation is associated both with the large molecular absorption in dense clouds, where protostars are localized, and with the absence of simultaneous monitoring data obtained using single radio telescopes and global interferometers. 
The most important issue is the localization of maser spots that cause such powerful short-lived flares of the water maser. 
Wherein, it is not at all necessary that the maser spots responsible for flares at the features near 6 and 8 km~$s^{-1}$ were in the same gas and dust formation, such as disks around protostars. 
In IRAS 16293 it is known that there are at least two such disks (A1, A2) with a distance of about 40 AU between them \cite{11}. 
These disks are of different sizes, and if maser spots are located on them, they will move at different radial velocities.  
In the case of A1 and A2 discs, these velocities do not exceed 20 and 9 km~$s^{-1}$. 
Line-of-sight velocities with respect LSR for them are 2.1 and 5.8 km~$s^{-1}$ respectively. 
We cannot indicate a variant of localization of maser spots, given that we do not have, interferometric data during these flare phenomena, however, we must, nevertheless, use the available interferometric data obtained in other epochs.

As already said, the maser spots responsible for flares at features near 6 and 8 km~$s^{-1}$ may be spatially unrelated to each other. 
Dzib and other in 2005-2006, conducted 15 interferometric sessions of IRAS 16293 observations in order to clarify the distance to the object by determining the trigonometric parallax of the water maser features. 
He found a single maser feature that was present in all 15-observation epochs, separated by a time of about 7 months. 
It was the strong maser emission at radial velocity $\approx$ 6.1 km~$s^{-1}$ \cite{34}. 
If this is the same spectral line as in our case, then it has not undergone significant changes in radial velocity. 
The resolution of interferometric observations was 0.2 km~$s^{-1}$. 
Dzib note that Imai and other \cite{22} also detected emission in their VERA observations from 2005 to 2006 at a velocity of 6.0 km~$s^{-1}$ with $S_v \approx$ 2 Jy, which may correspond to the emission at radial velocity $\approx$ 6.1 km~$s^{-1}$. 
Dzib found that the distance between the maser spots giving the lines near 6 km~$s^{-1}$, and A1, A2 is about 140 AU. 
In other words, the flares we detected at the features near 6 km~$s^{-1}$, are in no way connected with maser spots in disks around protostars. 
This is an unexpected and important conclusion.

Again, the question arises, where are these water masers located? 
The IRAS 16293 complex is a giant collapsing, rotating gas and dust cloud about 3000 AU in size. 
Its mass is about 4 $M_0$. 
The protostars, with which the disks A1 and A2 are connected, revolve around a common center of gravity with the indicated mass in Keplerian orbits \cite{11}. 
The A1 and A2 are separated by a distance of 40 AU immersed in a gas and dust cloud about 150 AU in size. 
The dust emission has a thermal spectrum close to the Planck one, with a luminosity of about 20 $L_0$. 
The masses of the components are respectively 0.8 and 1.4 $M_0$, the circulation period is 362 years. 
The system eccentricity within two rms is equal to zero. 
The obtained line-of-sight systematic velocities relative to the LSR for A1 and A2 are 2.1$\pm$0.1 km~$s^{-1}$, 5.8$\pm$0.1 km~$s^{-1}$ respectively. 
Gas-dust disks less than 3.6 AU and 12 AU in size are associated with A1 and A2 \cite{11}. 

Our data complements this picture with critical details. 
The emission from the water maser at the feature about 6 km~$s^{-1}$ consists of a set of emissions from 7 maser spots, possibly located in one or two maser cluster and been on the line of sight to the observer. 
The density of maser spots in the cluster is so high that the emissions of the maser spots partially overlap each other. 
We see this situation in Fig. 1. 
When passing from the emission of one spot to another, the velocities of the features change in the range of 0.2 km~$s^{-1}$ from 5.87 to 6.07 km~$s^{-1}$ (Table 3).  

This is completely new information that poses new questions for researchers. 
First, why the radiation near 6 km~$s^{-1}$ is preserved permanently at the same velocities for such a long time from 2005 to 2020? 
In addition, if this radiation is related to outflows, then how can this situation be explained. 
The question of where the water masers responsible for the powerful radiation at a velocity of about 6 km~$s^{-1}$ are located remains unanswered, given that there is no detectable source at the indicated position in the 3 mm microwave survey. 
This may means that our water masers also revolve around the common center of gravity, and being, in the first approximation, 140 AU from its center. 
Their location may be a protoplanetary gas and dust disk. 
Thus, we could possible find exactly the protoplanet (exoplanet) in the IRAS 16293 complex. 
In this situation, the protoplanet manifests itself only by a maser emission. 
In the same time, we cannot exclude an situation option by in which we do not see a low-mass protostar of solar mass near our maser spots, which is hidden from our eyes by a dense molecular cloud that has noticeable absorption even in the infrared range. 
For the case of young stars of small (solar) masses, Strelnitsky considered a model based on the idea of the formation of planetary systems around these young stars \cite{35}.

If there are no such protostars at all, then let us find the velocities of our masers in a circular Keplerian orbit around the general center of mass of 4 $M_0$. 
For a distance from the center of 140 AU, for example, we obtain $v$ = 5.1 km~$s^{-1}$. 
System A1, A2 has a regular velocity relative to LSR equal to $\approx$ 4 km~$s^{-1}$ \cite{11}. 
Taking into account uncertainties in the mass of the IRAS 16293 complex and the velocity of the general center of mass relative to the LSR, the radial velocities of the features can vary within 1-10 km~$s^{-1}$. 
Consequently, such a possibility of localization of the considered water masers cannot be ruled out.

Now we explain why we do not see a noticeable change in the radial velocities of features near 8 km~$s^{-1}$ over 15 years. 
The circulation period of our suppose system is about a thousand years. 
For 15 years, the system will rotate by only 1.5$\%$ of the total turnover. 
In principle, the magnitude of the change in the radial velocity of features with such rotation and higher resolution of interferometric observations could be fixed. 
However, this is worth considering in the future.
In the case of maser flares near 8 km~$s^{-1}$, we practically did not find any changes in the frequencies of the spectral lines of the flares on a time interval of 40 days at the level of observation errors ($\sim$0.1 km~$s^{-1}$). 
The question of the localization of these water masers remains open.

Another important question remains unclear: what initiates the maser emission in a gas and dust cloud remote from protostars? 
The variant of maser activation associated with the accretion of the surrounding matter onto a protostellar object is not preferable, since our features near 6 km~$s^{-1}$ are very far from A1 and A2. 
It remains to assume that our maser spots are located inside a dense massive core of gas and dust and move in a Keplerian orbit around a common center of mass. 

In addition to an outflow, initiator of maser emission can be a small-scale turbulence.  
Strong water masers may arise in a turbulent medium. 
This environment can see as a real source of collisional pumping of a water maser ("at least those corresponding to the low-velocity core of the spectrum") \cite{36,37,38}. 
H$_2$O masers can arise in star-forming regions due to regular movements of turbulent gas and dust matter during its expansion or accretion as it rotates  around protostars. 
A turbulence can cause powerful water masers, by providing energy for collisional pumping in small-scale stochastic shock waves.
The primary source of energy release in this case is the gravitational one due to the compressing gas and dust cloud of large dimensions 3000 AU \cite{11}. 
Strelnitsky modeled the considered method of collisional pumping of water masers using observational data of the known galactic sources W49N. W3(OH), Sergeant B2 (M). 
At the same time, an initiator of maser emission can be a small-scale turbulence \cite{37}. 

This allowed us to draw important conclusions that do not contradict our data: masers can exist in both saturated and unsaturated states when complex powerful flare phenomena arise.
In the saturated mode, a large number of spots with comparable brightness and a Gaussian turbulent line profile are observed. 
An initiator of maser emission can be a small-scale turbulence.  
A saturated maser has a lower flux density, than unsaturated ones. 
More numerous saturated masers can initiate the emission of several maser spots along the line of sight to the observer. 
For our case, this means that there is an increased probability of the saturated maser illuminating the unsaturated ones. 
For the water maser, at feature near 8 km~$s^{-1}$, the initiation of a powerful water maser Flare 3 by an ultra-short unsaturated Flare 1, providing an additional input flux for Flare 3 at about 2 kJy, is also detected. 
This indicates that a high input flux level of more than 0.5 kJy is required for the effective operation of the water maser when creating powerful flares. 
Another important detail: the measured by us line-width for saturated water masers towards IRAS 16293 is 2.8 km, in agreement with data \cite{39}. 
The absence of narrowing of the spectral line of the water maser during a flare is an important indication of the saturated state of the maser.

The unusual structure of the analyzed complex flare has the important common features with other flares, which we discovered in another galactic source W49N (at the feature -60 km~$s^{-1}$, \cite{26}). 
Similar to our case, powerful short flares occurred on top of a less powerful but more duration one. 
Its duration was approximately the same as that of the flare IRAS 16293 at the features near 6 and 8 km~$s^{-1}$. 
In all cases, it was also observed other main parameters of flares characteristic of an unsaturated maser - exponential increase and decrease in flux density, as well as the dependence of $\Delta v$ on $ln~S$. 
In all case, longer but less powerful flares were the starting points for powerful short flares. 

The new data we obtained for the phenomena considered of maser flares are very important.
Our data probably confirm the assumptions made at the early stages of the study of the water maser regarding the state of the maser in maser clusters \cite{40}. 
It is formulated as follows: a maser cluster can simultaneously contain maser formations in both the unsaturated and saturated states. 
When powerful flare phenomena occur, all participating in its creation flares are on the line of sight to the observer and have close velocities of features.  

Another important conclusion is that there may be maser spots responsible for powerful flares of the water maser, which are located in protoplanetary gas and dust formations and move in orbits around a common center of gravity, which includes not only masses associated with protostars.

\section{Conclusions}
\label{sec:con}

1. Detailed observations of the water maser of a protostellar object close to the solar mass IRAS 16293-2422 have been carried out for about two years.

2. Unusual powerful flare phenomena at features near 6 and 8 km~$s^{-1}$, consisting of individual water maser flares and emanating from maser spots located close to each other in maser clusters were discovered

3. Short power maser flares occurred at the top of a longer duration ones but less powerful flares, possibly initiating a powerful maser radiation of more short flares.
In a separate case, such a powerful initiator of maser emission was an ultra-short flare of an unsaturated water maser.

4. It was also established for the emergence of powerful water masers, a sufficiently high (more than 0.5 kJy) input flux into the maser formation is required.

5. Important physical parameters of IRAS 16293-2224 water maser flares have been obtained.

6. The possible localization of water masers at features near 6 and 8 km~$s^{-1}$ within the young gas and dust complex IRAS 16293-2224 has been established.

\acknowledgments

The RT-22 observations, collecting and data processing were done under support of the Ministry of Science and Higher Education of the Russian Federation under the grant 075-15-2020-780. We wish to thank the Maser Monitoring Organisation (M2O).  


\end{document}